\documentclass[onecolumn,aps,floatfix,pra,superscriptaddress,nofootinbib]{revtex4}

\usepackage{color}
\usepackage{graphicx}
\usepackage{bm}
\usepackage{amsmath}
\usepackage{hyperref}

\usepackage{upgreek}
\usepackage[subnum]{cases}

\newcommand{\pa}{ \partial }
\newcommand{\hb}{ \hbar }
\newcommand{\si}{ \sigma }

\newcommand{\ga}{ \gamma }
\newcommand{\la}{ \langle }
\newcommand{\ra}{ \rangle }
\newcommand{\del}{ \delta }

\newcommand{\re}{ \text{Re} }

\newcommand{\sip}{ \text{sp} }

\newcommand{\MB}{ \text{MB} }
\newcommand{\BE}{ \text{BE} }
\newcommand{\FD}{ \text{FD} }

\begin{document}

\title{Momentum-space decoherence of distinguishable and identical particles  in the Caldeira-Leggett formalism}

\author{Z. Khani}
\email{z.khani@stu.qom.ac.ir}
\affiliation{Department of Physics, University of Qom, Ghadir Blvd., Qom 371614-6611, Iran}
\author{S. V. Mousavi}
\email{vmousavi@qom.ac.ir}
\affiliation{Department of Physics, University of Qom, Ghadir Blvd., Qom 371614-6611, Iran}
\author{S. Miret-Art\'es}
\email{s.miret@iff.csic.es}
\affiliation{Instituto de F\'isica Fundamental, Consejo Superior de Investigaciones Cient\'ificas, Serrano 123, 28006 Madrid, Spain}
\begin{abstract}
In this work, momentum-space decoherence using minimum and nonminimum-uncertainty-product (stretched) Gaussian wave packets in the framework of Caldeira-Leggett formalism and under the presence of a linear potential is studied. 
As a dimensionless measure of decoherence, purity, a quantity appearing in the definition of the {\it linear entropy}, is studied taking into account the role of the stretching parameter. 
Special emphasis is on the open dynamics of the well-known cat states and bosons and fermions compared to distinguishable particles.
For the cat state, while the stretching parameter speeds up the decoherence, the external linear potential strength does not affect the decoherence time; 
only the interference pattern is shifted. Furthermore, the interference pattern is not observed 
for minimum-uncertainty-product-Gaussian wave packets in the momentum space. 
Concerning bosons and fermions, the question we have  addressed is how  the symmetry of the wave functions of indistinguishable particles 
is manifested in the decoherence process, which is understood here as the loss of being indistinguishable due to 
the gradual emergence of classical statistics with time. We have observed that the initial  bunching and anti-bunching character of bosons and 
fermions, respectively, in the momentum space are not preserved as a function of the environmental parameters, temperature and damping constant.  
However, fermionic distributions are slightly broader than the distinguishable ones and these similar to the bosonic distributions.
This general behavior could be interpreted as a residual reminder of the symmetry of the wave functions
in the momentum space for this open dynamics.

\end{abstract}

\maketitle

{\bf keyword}:
Decoherence; Caldeira-Leggett formalism; Momentum space; Stretched Gaussian wave packet; Cat state; Bosons; Fermions

%%%%%%%%%%%%%%%%%%%%%%%%%%%%%%%%%%%%%%%%%%

\section{Introduction}

Decoherence is a crucial process in order to better understand the emergence of classical behavior in the quantum dynamics of physical systems
\cite{Zeh,Zurek,Joos}.
This process arises when the physical system of interest interacts with an apparatus to carry out a measurement or when it is immersed in a
given environment. The theory of open quantum systems is the natural framework to carry out this kind of studies and has been widely developed 
from quite different approaches and published in several books \cite{Percival,Weiss,Mensky,BaPe-2002,Joos-2003,Sch-2007,Caldeira-2014,Salva}. 
Within the theoretical methods working with wave functions instead of reduced density matrix, one can find some 
approaches  within the so-called Caldirola-Kanai and Scr\"odinger-Langevin frameworks \cite{MoMi-AP-2018, MoMi-JPC-2018, MoMi-EPJP-2019-1, MoMi-EPJP-2019-2}. Both approaches are not following 
the system-plus-environment model but effective time dependent  Hamiltonians and nonlinear Schr\"odinger equations, respectively. 
Recently, interference and diffraction of identical spinless particles in one slit problems \cite{MoMi-EPJP-2020-1} have been analyzed.

In this work, we are going to focus on the so-called Caldeira-Leggett (CL) formalism \cite{CaLe-PA-1983,Caldeira-2014}. This formalism is based on
the reduced density matrix once one carries out the integration over the environmental degrees of freedom. As is well-known, the diagonal matrix
elements give probabilities and off-diagonal matrix elements are called coherences. In the decoherence process these off-diagonal elements go to 
zero more or less rapidly depending on the parameters characterizing the environment; usually, damping constant and temperature. Most of
studies involving quantum decoherence is being carried out in the configuration space and very few in the momentum space.
Venugopalan \cite{Ve-PRA-1994} studied the decoherence of a {\it free single} minimum-uncertainty-product Gaussian wavepacket in the CL 
formalism within the context of measurement processes both in position and momentum spaces. This study  revealed that the emergent 
{\it preferred basis} selected by the environment is the momentum basis.
By considering a cat state, decoherence {\it without dissipation} has been studied in phase space \cite{FC-Arxiv-2003}. To this end, 
these authors considered the quantum system in thermal equilibrium and assumed a  weak interaction with the environment in a way that dissipation 
could be neglected. Then, from principles of statistical mechanics, the corresponding probability distribution were obtained by averaging over a 
thermal distribution of velocities. Furthermore, the Wigner phase space distribution function was also obtained and the destruction of the 
interference term was studied as a function of time.  Decoherence was claimed not to occur in momentum and phase space. 
More recently, decoherence in momentum space has been studied in the context of suppression of quantum-mechanical reflection 
\cite{BeHa-PRA-2013} using a master equation resembling the CL equation \cite{CaLe-PA-1983, Caldeira-2014} in the negligible dissipation limit; and for a non-relativistic charged particle described by a wave packet under the presence of linear interaction with the electromagnetic field in equilibrium at a certain temperature \cite{BeCoPe-AIP-2004}.
Recently, in the chemical physics community,  studies about purity are also found  questioning this quantity as a measure of decoherence in the dynamics of quantum dissipative systms \cite{Makri,Franco, Singh}.

The central goal of this work is to show how decoherence affects the open dynamics of cat states and identical spinless (bosons and fermions) 
particles within the momentum representation, far less investigated than in the configuration space \cite{MoMi-submitted}. For cat states, while the stretching parameter speeds 
up the decoherence, the external linear potential strength does not affect the decoherence time; only the interference pattern is shifted. Furthermore, 
the interference pattern is not observed for minimum-uncertainty-product-Gaussian wave packets in the momentum space.
Purity, a quantity appeared in the definition of the {\it linear entropy}, and its relation to coherence length is studied taking into account the role of the stretching parameter. 
The next question is how  the symmetry of the corresponding wave functions is manifested in the decoherence process. 
This process is understood here as the loss of the indistinguishable character of those particles due to 
the gradual emergence of classical statistics with time. In particular, the well-known bunching and anti-bunching properties of bosons and 
fermions, respectively, when  minimum and non-minimum-uncertainty-product Gaussian wavepackets are used is considered  as a function of the environmental 
parameters, temperature and damping constant. We have observed that the symmetry of the initial distribution 
is not preserved in the time evolution of the corresponding wave functions. However, fermionic distributions are slightly
broader than the distinguishable ones and these are similar to the corresponding bosonic distributions.
This could be interpreted as a residual reminder of the bunching and anti-bunching character of the initial distributions in the momentum space
but washing them out when increasing the damping constant and temperature.
This general behavior has also been confirmed by carrying out a different theoretical analysis from the single-particle probability. Finally, an indirect 
manifestation of these properties for bosons and fermions have also been observed when considering the so-called simultaneous detection probability

This paper is organized as follows. In Section II the CL master equation in the momentum representation is briefly introduced. In Section III, 
open dynamics and decoherence of minimum and non-minimum-uncertainty-product Gaussian wavepackets are analyzed for cat states and 
under the presence  of a linear potential.  Then, open dynamics of two identical spinless particles (bosons and fermions) are analyzed in Section IV.  
In Section V, results, discussion and some concluding remarks are presented.

\section{The Caldeira-Leggett master equation in the momentum representation}
\label{sec: CL}

In the context of open quantum systems and considering the reservoir as a set of non-interacting oscillators, Caldeira and Leggett obtained the well-known master equation \cite{CaLe-PA-1983, Caldeira-2014}
\begin{eqnarray} \label{eq: CLeq_repind}
\frac{\pa \hat{\rho}}{\pa t} &=& 
\frac{1}{i\hb} [\hat{H}_0, \hat{\rho}] + \frac{\ga}{i\hb} [ \hat{x}, \{\hat{p}, \hat{\rho}\}] - \frac{D}{\hb^2}
[\hat{x}, [\hat{x}, \hat{\rho}]]  
\end{eqnarray}
for the reduced density matrix of the system
%, where $\hat{\cdot}$ over quantities mean operators, 
where $\ga$ is the damping constant or dissipation rate 
and $D = 2m\ga k_B T$ plays the role of the diffusion coefficient with $m$ the mass of particles; $k_B$ and $T$ being Boltzmann's 
constant and the environment temperature, respectively. The Hamiltonian $\hat{H}_0$ is given by
\begin{eqnarray} 
\hat{H}_0 &=& \frac{\hat{p}^2}{2m} + \hat{V}  .
\end{eqnarray}
Eq. (\ref{eq: CLeq_repind}) in the momentum representation for an external potential $ \hat{V} = V(\hat{x}, \hat{p}) $ reads as
\begin{eqnarray} \label{eq: CLeq_mom}
\frac{\pa}{\pa t} \rho(p, p', t) &=& 
\bigg[-\frac{i}{2 m \hb} (p^2 -p'^2) + \frac{ V\left( i\hb \frac{\pa}{\pa p}, p \right) - V\left( -i\hb \frac{\pa}{\pa p'}, p' \right) }{i\hb} 
\nonumber\\ 
& & +
 \ga \left( \frac{\pa}{\pa p} + \frac{\pa}{\pa p'} \right)(p+p') 
+ D \left( \frac{\pa}{\pa p} + \frac{\pa}{\pa p'} \right)^2 \bigg] \rho(p, p', t) 
\end{eqnarray}
where the off-diagonal matrix elements are $ \rho(p, p', t) = \la p | \hat{\rho} | p' \ra$ and known as coherences. 
In the center of mass and relative coordinates
\begin{numcases}~
u = \frac{p+p'}{2} \label{eq: cm} \\
v = p - p'  \label{eq: rel} 
\end{numcases}
Eq. (\ref{eq: CLeq_mom}) for the external linear potential $ \hat{V} = m g \hat{x} $ can be expressed as
\begin{eqnarray} \label{eq: CLeq_uv}
\frac{\pa }{\pa t} \rho(u, v, t) + \frac{\pa }{\pa u} j(u, v, t) + \frac{i}{m\hb} u v \rho(u, v, t) &=& 0  .
\end{eqnarray}
the current density matrix being
\begin{eqnarray} \label{eq: cur_den_mat}
j(u, v, t) &=& - \left( m g + 2 \ga u  + D \frac{\pa}{\pa u} \right) \rho(u, v, t) .
\end{eqnarray}
As is known, when $v=0$, the diagonal elements of the density matrix gives the probability density and the continuity equation is written as
\begin{eqnarray} \label{eq: con_CL}
\frac{\pa P(p, t)}{\pa t} + \frac{\pa J(p, t)}{\pa p}  &=& 0 ,
\end{eqnarray}
where $P(p, t)$ and $J(p, t)$ are the diagonal elements of $\rho(u,v,t)$ and $j(u,v,t)$, respectively. 

%One can then construct Bohmian trajectories $ p(p^{(0)}, t) $ in the momentum representation \cite{BoSc-PLA-2020} by solving the guidance equation
%
%\begin{eqnarray} \label{eq: guidance}
%\dot{p}(p, t) = \frac{J(p, t)}{P(p, t)}\bigg|_{p = p(p^{(0)}, t)} ;
%\end{eqnarray}
%
%$ p^{(0)} $ being the initial momentum of the Bohmian particle.

\section{Open dynamics and decoherence of Gaussian wave packets. The cat state.}

Let us consider a linear potential given by $\hat{V} = m g \hat{x}$ for nonminimum-uncertainty-product or {\it stretched} Gaussian wave packets 
in the CL framework for two cases:  the open dynamics of a single wave packet  and afterwards the corresponding dynamics for a pure initial state 
consisting of superposition of two well separated wavepackets, a cat state.

\subsection{A single Gaussian wave packet in a linear potential}

For a single Gaussian wave packet, the solution of Eq. (\ref{eq: CLeq_uv}) can be easily found by assuming the Gaussian ansatz,
\begin{eqnarray} \label{eq: rho_G}
\rho(u, v, t) &=& \frac{1}{\sqrt{2\pi d_2(t)}} \exp \left[ d_0(v, t) - \frac{ (u-d_1(v, t))^2 }{ 4d_2(t) } \right]
\end{eqnarray}
and from Eq. (\ref{eq: cur_den_mat}), one has that
\begin{eqnarray} \label{eq: j_G}
j(u, v, t) &=& \left[ - m g -2\ga u + \frac{D}{2d_2(t)}(u-d_1(v, t)) \right] \rho(u, v, t) .
\end{eqnarray}

On the other hand, let us consider the initial state as the stretched Gaussian wave packet whose Fourier transform takes the form 
\begin{eqnarray} \label{eq: wf0}
\phi_0(p) &=& \left( \frac{2}{\pi} \frac{\si_0^2}{\hb^2} \right)^{1/4} \exp \left[ - (1+i\eta) \frac{ (p-p_0)^2 \si_0^2}{\hb^2} - i \frac{(p- p_0) x_0}{\hb} \right]   .
\end{eqnarray}
Here, $x_0$ and $p_0$ are the center and kick momentum and $\eta$ is the stretching parameter 
governing the position-space width, $ \Delta x = \si_0 \sqrt{1 + \eta^2}$. Thus, the uncertainty product $ \Delta x \Delta p 
= \frac{\hb}{2} \sqrt{1 + \eta^2} $ reaches the minimum value for $\eta=0$. With this in mind, the solution of Eq. (\ref{eq: rho_G}) reads

\begin{numcases}~
d_0(v, t) = - \frac{i}{\hb} x_t ~ v - \left[ (\eta^2+1) \frac{\si_0^2}{2\hb^2} + \eta \frac{\uptau(t)}{2m\hb} + \frac{ \uptau(t)^2 }{8 m^2 \si_0^2} - D \frac{ 3 + e^{-4\ga t} - 4 e^{-2\ga t} - 4\ga t }{ 16 \hb^2 m^2 \ga^3}  
\right] v^2
\label{eq: d0} \\
d_1(v, t) = p_t - i \left[ \left( \frac{\hb}{4m\si_0^2}  \uptau(t) + \frac{\eta}{2} \right) e^{-2\ga t} + \frac{D}{m\hb} \uptau(t)^2  \right] v 
\label{eq: d1} \\ 
d_2(t) = \frac{\hb^2}{8\si_0^2} e^{-4\ga t} + D \frac{1-e^{-4\ga t}}{4\ga}
\label{eq: d2}
\end{numcases}
with
\begin{numcases}~ 
x_t = x_0 + \frac{p_0}{m} \uptau(t) + g \frac{\uptau(t)-t}{2\ga} \label{eq: xt} \\
p_t = p_0 e^{-2\ga t} - m g \uptau(t) \label{eq: pt} \\
\uptau(t) = \frac{1-e^{-2\ga t}}{2\ga}     .   \label{eq: tau}
\end{numcases}
Note that $ x_t $ is the trajectory followed by a {\it classical} particle with mass $m$ and initial velocity $p_0/m$ immerse in a viscid media 
with a damping constant $\ga$ and under the presence of a constant force field $-mg$; and $ p_t = m \dot{x}_t$ \cite{MoMi-JPC-2018}.
By imposing the conditions $v=0$, the probability density (PD) and the probability current density (PCD) are expressed as
\begin{eqnarray}
P(p, t)& =& \frac{1}{ \sqrt{2\pi} w_t } \exp\left[ - \frac{ ( p - p_t )^2 }{ 2 w_t^2} \right] \label{eq: probden}  \\
J(p, t) &=& \left[ -m g -2\ga p + \frac{D}{w_t^2}( p - p_t ) \right] P(p, t)
\label{eq: cur}
\end{eqnarray}
with
\begin{eqnarray} 
w_t &=& \sqrt{2 d_2(t)} = e^{-2\ga t} \frac{\hb}{2\si_0} \sqrt{ 1 + D \frac{2\si_0^2}{\hb^2 \ga } (e^{4\ga t}-1) } \label{eq: wt}
\end{eqnarray}
being the width of the distribution function in momentum space.
As Eqs. (\ref{eq: probden}) and (\ref{eq: cur}) clearly show, in the momentum representation, the stretching parameter $\eta$ plays no role in the  
PD and PCD.  
%

%\subsubsection{Short time limit and decoherence time}

%\subsubsection{Long time limit and preferred basis of measurement}

In the long time limit, $\ga t \gg 1$, only terms which are constant and/or depend linearly on time survive and one has that
\begin{numcases}~
x_t  \approx  - \frac{g}{2\ga} t ,\\
p_t \approx - \frac{mg}{2\ga} ,\\ 
d_0(t) \approx  \left( \frac{i}{\hb} \frac{g}{2\ga} v - \frac{D}{ 4 \hb^2 m^2 \ga^2} v^2 \right)t , \\
d_1(t) \approx - \frac{mg}{2\ga} - i \frac{D}{4m \hb \ga^2} v , \\
d_2(t) \approx \frac{D}{4\ga}, 
\end{numcases}
yielding
\begin{eqnarray} \label{eq: rho_long}
\rho(u, v, t) & \approx & \sqrt{ \frac{\ga}{\pi D} } \exp \left[  
- \frac{4\ga^2 u^2 + 4 m \ga g u + m^2 g^2}{4\ga D} + 
\left( \frac{i}{\hb} \frac{g}{2\ga} v - \frac{D}{4  \hb^2 m^2 \ga^2} v^2 \right) t
\right]
\end{eqnarray}
showing that the off-diagonal elements of the reduced density matrix, $v \neq 0 $, decay exponentially with  time. This allows us to define
a time as
\begin{eqnarray} \label{eq: td}
t_d &=& \frac{ 4 \hb^2 m^2 \ga^2 }{ D v^2 } ,
\end{eqnarray}
which is the characterisitc time required to damp momentum coherences over a distance $v$. The inverse $t_d^{-1}$ plays the role of 
a {\it decoherence} rate. Thus, the momentum space is the obvious choice for the preferred basis as already mentioned in \cite{Ve-PRA-1994}.

\subsection{Purity and coherence length in momentum space}

\begin{figure} 
\centering
\includegraphics[width=12cm,angle=-0]{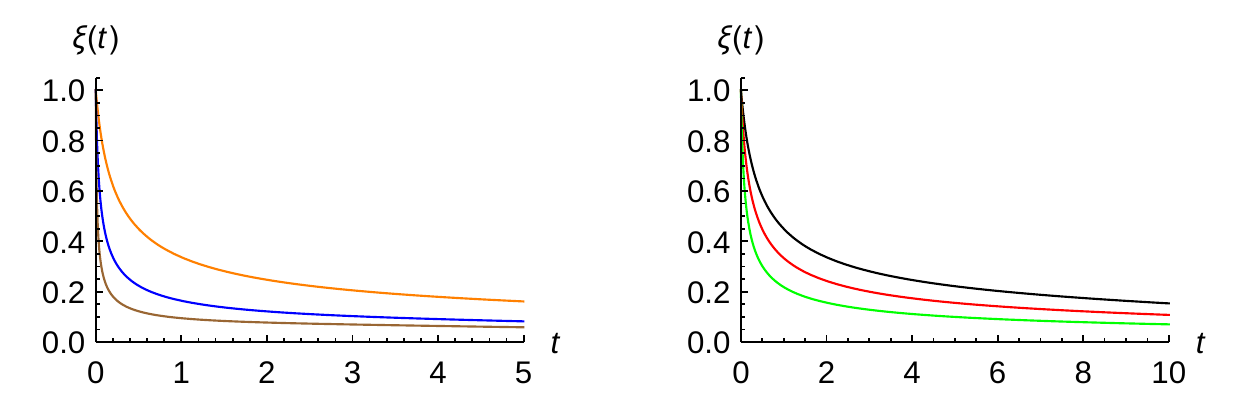}
\caption{ Purity $ \xi(t) $ given by Eq. (\ref{eq: pu_G}) for $\si_0=5$ and $ k_B T = 2 $; and for minimum-uncertainty-Gaussian wavepacket with different values of the damping constant (left panel) and for $\ga =0.005$ but with different values of the stretching parameter (right panel). 
Color curve codes are in the left panel: $ \ga = 0.01 $ (orange), $ \ga = 0.05 $ (blue), $ \ga = 0.2 $ (brown); whereas in the right panel:
$ \eta = 0 $ (black), $ \eta = 1 $ (red) and $ \eta = 2 $ (green).}
	\label{fig: pu} 
\end{figure}

As is known, a pure state can not be preserved along its open dynamics. This can be easily seen by evaluating the trace of the square of density matrix, $ \hat{\rho}^2(t) $, or purity 
\begin{eqnarray} \label{eq: pu}
\xi(t) &=& \int dp \int dp' | \la p | \hat{\rho} | p' \ra |^2 
= \int du \int dv | \rho(u, v, t) |^2 .
\end{eqnarray}
Writing Eqs. (\ref{eq: d0}) and (\ref{eq: d1}) in the form
\begin{numcases}~
d_0(v, t) = - d_{02}(t)~ v^2 - i ~ d_{01}(t) ~ v  \label{d0_new}\\
d_1(v, t) = - d_{10}(t) - i ~ d_{11}(t) ~ v \label{d1_new}
\end{numcases}
where the new  coefficients are very easily identified, one obtains 
\begin{eqnarray} \label{eq: pu_G}
\xi(t) &=& \frac{1}{ 2\sqrt{ 4 ~ d_2(t) ~ d_{02}(t) - ( d_{11}(t) )^2 } }
\end{eqnarray}
for the Gaussian solution given by Eq. (\ref{eq: rho_G}). Then, from Eqs. (\ref{eq: d0}), (\ref{eq: d1}) and (\ref{eq: d2})  the purity $ \xi(t) $ is 
an independent quantity on $p_0$, $x_0$ and the field strength $g$. As expected,  it becomes unity for $\ga = 0 $. 
Expanding $ \xi(t) $ in powers of $ t $ yields
\begin{eqnarray} \label{eq: pu_G-approx}
\xi(t) &=& 1 + \left[ 2 \ga - \frac{4\si_0^2(1+\eta^2)}{\hb^2}  D \right]t + O(t^2)
\end{eqnarray}
whereas in the long time limit, $ \ga t \gg 1 $, purity becomes zero. 
We should stress that this linear behaviour in short times is a typical feature of master Markovian regime \cite{BaPe-2002, Joos-2003}. In spite of this result, there are proofs in favour of the quadratic behaviour \cite{all-PRL-1996, GuFr-JPCL-2017}. However, the point is that to make the proof, invariance of the trace under cyclic permutation has to be used which is questionable when the state space is infinite-dimensional. There is another proof in favour of the quadratic behaviour where the researchers consider time evolution of the Wigner function in phase space using a von Neumann-like equation but with the Moyal bracket of Hamiltonian and the Wigner function instead of their commutator \cite{GoBr-PRL-2003}. 
As an illustration, in Fig. \ref{fig: pu} the evolution of purity with time is 
plotted for different damping constants (left panel) and stretching parameters (right panel) for $\si_0=5$ and $ k_B T = 2 $. This quantity 
decays faster with $\ga$ than with $\eta$. The same behavior is expected when increasing the temperature although it is not shown in this figure.

The range of spatial coherence in momentum space can also be quantified by the off-diagonal direction $ p = - p'$ \cite{Sch-2007}. 
From Eqs. (\ref{eq: rho_G}), (\ref{d0_new}) and (\ref{d1_new}) one observes that the width of the Gaussian in this direction is
\begin{eqnarray} \label{eq: cl_G}
\mu(t) &=& \sqrt{ \frac{ d_2(t) }{ 2[ 4 d_2(t) ~ d_{02}(t) - ( d_{11}(t) )^2 ] } }
\end{eqnarray}
which can be interpreted as the {\it coherence length in momentum space} \cite{Joos-2003}. Interestingly enough, the ratio of this coherence length 
and the distribution width provides again  the purity
\begin{eqnarray} \label{eq: pucl_G}
\frac{ \mu(t) }{ w_t } &=& \xi(t)
\end{eqnarray}
where we have used Eq. (\ref{eq: pu_G}). This reveals that purity $ \xi(t) $ can also be interpreted as a dimensionless measure of decoherence \cite{Joos-2003}.
From Eqs. (\ref{eq: d0}), (\ref{eq: d1}) and (\ref{eq: d2}) one sees that $ \mu(t) $ decreases with the stretching parameter $\eta$. By expanding up to the second order in $ t $, one obtains
\begin{eqnarray} \label{eq: cl_G-short}
\mu(t) & \simeq & \frac{\hb}{2\si_0} \left\{ 1 - 4 \eta^2 \frac{\si_0^2 D}{\hb^2} t + 
D \left[ - \frac{2}{m\hb} \eta + \frac{8 \si_0^2 \ga }{ \hb^2 }\left( -\ga + \frac{ \si_0^2 }{ \hb^2 } (4+3\eta^2) D \right) \eta^2  \right] t^2 \right \}.
\end{eqnarray}
Note that for minimum-uncertainty-product wavepackets i.e., $ \eta=0 $, there are no linear and square terms in time. 
At long times, the coherence length vanishes according to
\begin{eqnarray} \label{eq: cl_G-long}
\mu(t) & \simeq & 2 m \ga \hb \sqrt{ \frac{2}{D ~ t} }.
\end{eqnarray}

\subsection{The cat state}

Let us consider now the initial state as a superposition of two well separated wave packets in momentum space,
\begin{eqnarray} \label{eq: sup0}
\phi_0(p) &=& \mathcal{N} ( \phi_{0a}(p) + \phi_{0b}(p) ) 
\end{eqnarray}
$ \mathcal{N} $ being the normalization constant.
From Eq. (\ref{eq: sup0}), the initial density matrix has the form
\begin{eqnarray} \label{eq: rho0}
\rho(p, p', 0) &=& \mathcal{N}^2 ( \rho_{aa}(p, p', 0) + \rho_{ab}(p, p', 0) + \rho_{ba}(p, p', 0) + \rho_{bb}(p, p', 0) ) 
\end{eqnarray}
where $ \rho_{ij}(p, p', 0) = \phi_{0i}(p) \phi_{0j}^*(p') $; $i$ and $j$ being $a$ or $b$. Due to the linearity of the master equation (\ref{eq: CLeq_mom}), one obtains again the evolution of each term of 
Eq. (\ref{eq: rho0}) separately by using the method outlined above i.e., by assuming a Gaussian ansatz. 
Afterwards, these solutions are superposed to have the time 
dependent  PD  according to \cite{MoMi-EPJP-2020-1}, 
\begin{eqnarray} \label{eq: probden_sup}
	P(p, t) &=& \mathcal{N}^2 ( P_{aa}(p, t) + P_{ab}(p, t) + P_{ba}(p, t) + P_{bb}(p, t) )   .
\end{eqnarray}
By using the fact that $ P_{ba}(p, t) = P_{ab}^*(p, t) $, one can write
\begin{eqnarray} \label{eq: probden_sup1}
P(p, t) &=& \mathcal{N}^2 ( P_{aa}(p, t) + P_{bb}(p, t) + 2 |P_{ab}(p, t)| \cos \Theta(p, t) ) 
\end{eqnarray}
where $|P_{ab}(p, t)|$ is the modulus of $P_{ab}(p, t)$ and $ \Theta(p, t) $  its phase. Rewriting Eq. (\ref{eq: probden_sup1}) as the 
typical interference pattern expression \cite{BaPe-2002}
\begin{eqnarray} \label{eq: probden_sup2}
P(p, t) &=& \mathcal{N}^2 ( P_{aa}(p, t) + P_{bb}(p, t) + 2 \sqrt{ P_{aa}(p, t) P_{bb}(p, t) } ~ e^{\Gamma(t)} 
\cos \Theta(p, t) ) 
\end{eqnarray}
one has that
\begin{eqnarray} 
\Gamma(t) &=& \log \frac{|P_{ab}(p, t)|}{ \sqrt{ P_{aa}(p, t) P_{bb}(p, t) } }    ,
\end{eqnarray}
$\Gamma(t)$ being the so-called decoherence  function which is negative. The  corresponding exponential function 
\begin{eqnarray} \label{eq: atten}
a(t) &=& e^{\Gamma (t)}
\end{eqnarray} 
is called the coherence attenuation coefficient which quantifies the reduction of the interference visibility \cite{FC-PLA-2001}. 

Let us assume that the two  wavepackets $ \phi_{0a}(p) $ and $ \phi_{0b}(p) $ are stretched Gaussian functions,  Eq. (\ref{eq: wf0}), 
co-centered in position space, $ x_{0a} = x_{0b} = 0 $,  having the same stretching parameter $\eta$, width $\si_0$ and different kick 
momenta, $p_{0a}$ and $p_{0b}$. 

\subsubsection{Free evolution}

Then, the evolution of the cross term $ \rho_{ab}(p, p', 0) = \phi_{0a}(p) \phi_{0b}^*(p') $ is given by the Gaussian ansatz (\ref{eq: rho_G}) with 
\begin{eqnarray}
d_{0,ab}(v, t)& =& - \frac{ ( p_{0a} - p_{0b} )^2 \si_0^2 }{ 2\hb^2 }(1+\eta^2)  
+ \left[ \frac{( p_{0a} - p_{0b} ) \si_0^2}{\hb^2}(1+\eta^2) - i \frac{\uptau(t)}{ 2\hb m}[( p_{0a} + p_{0b} )+i( p_{0a} - p_{0b} )\eta] \right] v
\nonumber \\
& &- \left[ 
\frac{\si_0^2}{2\hb^2}(1+\eta^2) - \frac{ \uptau(t)^2 }{8m^2 \si_0^2}  + \frac{\uptau(t)}{ 2\hb m} \eta
- D \frac{ 3 + e^{-4\ga t} - 4 e^{-2\ga t} - 4\ga t }{ 16 \hb^2 m^2 \ga^3}  
\right] v^2
\label{eq: d0abf}
	\\
d_{1,ab}(v, t) &=& \frac{1}{2} e^{-2\ga t} [( p_{0a} + p_{0b} )+i( p_{0a} - p_{0b} )\eta] - i \left[ \frac{ \uptau(t) }{ m } \left( \frac{\hb}{4\si_0^2} e^{-2\ga t} + \frac{D}{\hb} \uptau(t)  \right) + \frac{1}{2} e^{-2\ga t} \eta \right] v 
\label{eq: d1abf}
\end{eqnarray}
Note that the additional subscript $ab$ refers to the cross term $ \rho_{ab}(p, p', 0)$. For the evolution of the remaining terms of (\ref{eq: rho0}), 
one just uses the corresponding momenta in Eqs. (\ref{eq: d0abf}) and (\ref{eq: d1abf});
the function $ d_2(t) $ remains the same as Eq. (\ref{eq: d2}).

\subsubsection{Linear potential}

In the presence of the external linear potential $ \hat{V} = m g \hat{x} $, the evolution of the cross term $ \rho_{ab}(p, p', 0) = \phi_{0a}(p) \phi_{0b}^*(p') $ is given by the same Gaussian ansatz (\ref{eq: rho_G}) with 
\begin{eqnarray}
d_{0,ab}(v, t)& =& - \frac{ ( p_{0a} - p_{0b} )^2 \si_0^2 }{ 2\hb^2 }(1+\eta^2)  
+ \left[ \frac{( p_{0a} - p_{0b} ) \si_0^2}{\hb^2} - i \frac{\uptau(t)}{ 2\hb m}[( p_{0a} + p_{0b} )+i( p_{0a} - p_{0b} )\eta] + i \frac{ t - \uptau(t) }{ 2h\ga }g \right] v
\nonumber \\
& &- \left[ 
\frac{\si_0^2}{2\hb^2}(1+\eta^2) - \frac{ \uptau(t)^2 }{8m^2 \si_0^2}  + \frac{\uptau(t)}{ 2\hb m} \eta
- D \frac{ 3 + e^{-4\ga t} - 4 e^{-2\ga t} - 4\ga t }{ 16 \hb^2 m^2 \ga^3}  
\right] v^2
\label{eq: d0abl}
	\\
d_{1,ab}(v, t) &=& \frac{1}{2} e^{-2\ga t} [( p_{0a} + p_{0b} )+i( p_{0a} - p_{0b} )\eta]  - m g \uptau(t) - i \left[ \frac{ \uptau(t) }{ m } \left( \frac{\hb}{4\si_0^2} e^{-2\ga t} + \frac{D}{\hb} \uptau(t)  \right) + \frac{1}{2} e^{-2\ga t} \eta \right] v   .
\label{eq: d1abl}
\end{eqnarray}
Analogously, for the evolution of the remaining terms of Eq. (\ref{eq: rho0}), one just uses the corresponding momenta in Eqs. (\ref{eq: d0abl}) 
and (\ref{eq: d1abl}).  Again, the function $ d_2(t) $ is given by the same  Eq. (\ref{eq: d2}).

\subsection{Decoherence}

If the initial state is now a superposition of two stretched Gaussian wave packets with the same width and located symmetrically around the origin
of momenta
\begin{eqnarray} 
\phi_0(p) &=& \mathcal{N} \left( \frac{ 2\si_0^2 }{ \pi \hb^2 } \right)^{1/4} \left\{
\exp \left[ - ( 1 + i \eta ) \frac{ ( p-p_0 )^2 \si_0^2 }{\hb^2} \right]
+ \exp \left[ - ( 1 + i \eta ) \frac{ ( p + p_0 )^2 \si_0^2 }{\hb^2} \right]
 \right\} 
 \nonumber \\
 \label{eq: wf0_sup_non} 
\end{eqnarray}
where the normalization constant $ \mathcal{N} $ is 
\begin{eqnarray} \label{eq: nor}
\mathcal{N} &=& \left\{ 2 + 2 \exp \left[ - \frac{2 p_0^2 (1+\eta^2) \si_0^2}{\hb^2}  \right] \right\}^{-1/2}  
\end{eqnarray}
one readily obtains
\begin{numcases}~
\Gamma(t) = - \frac{ 8 p_0^2 \si_0^4 ( 1 + \eta^2 ) }{ \hb^2 } \frac{  \sinh(2\ga t) }{ \hb^2 \ga e^{-2\ga t} + 4 D \si_0^2 \sinh(2\ga t) } D \label{eq: deco_func}
\\
\Theta(p, t) = 4 p_0 \ga \eta \si_0^2 \frac{ p + mg \uptau(t) }{ \hb^2 \ga e^{-2\ga t} + 4 D \si_0^2 \sinh(2\ga t) } \label{eq: phase}
\end{numcases}
for the decoherence function and phase, respectively. 
Eq. (\ref{eq: deco_func}) shows that $ \Gamma(t) = 0 $ for $D=0$ implying that the last term in Eq. (\ref{eq: CLeq_mom}) is responsible 
for decoherence \cite{Zu-PT-1991}. The stretching parameter $\eta$ speeds up the decoherence process. The external linear force does not affect 
the decoherence process; only the interference pattern is shifted. Furthermore,
from Eq. (\ref{eq: phase}), it is apparent that the phase function is zero for $\eta=0$ i.e., the interference pattern is not observed 
for minimum-uncertainty-product-Gaussian wave packets. This behavior is expected to also occur in isolated systems obeying the Schr\"{o}dinger equation,
\begin{eqnarray}
\phi(p, t) &=& \la p | \phi(t) \ra = \la p | \hat{U}(t) | \phi(0) \ra = e^{-i p^2 t /(2m\hb)} \phi(p, 0)
\end{eqnarray}
where for simplicity we have considered free propagation. There is only an overall phase factor. Thus, one has $ |\phi(p, t)| = |\phi(p, 0)| $ and 
from which $ \Theta(p, t) = \eta \dfrac{4 p_0 \si_0^2}{\hb^2} p $; since the two wavepackets are well separated in the $p-$space with no overlapping, 
the interference term is practically zero, $ P_{aa} P_{bb} \simeq 0$.

In the long times limit, the decoherence function reaches the asymptotic value
\begin{eqnarray} \label{eq: Ga_infinity}
\Gamma_{\infty} & \approx & - \frac{p_0^2}{2\si_p^2}
\end{eqnarray}
where $ \si_p = \hb / 2\si_0 $. %is the initial width of the two Gaussian functions. 
%$ \Gamma_{\infty} $ is proportional to the square of the ratio of initial distance between component waves, $2p_0$, to the initial width.  
%
In the negligible dissipation limit where the third term in the right hand side of Eq. (\ref{eq: CLeq_mom}) is neglected, one has that
\begin{numcases}~
\Gamma(t) \approx - \frac{ 16 p_0^2 \si_0^4 ( 1 + \eta^2 ) D }{ \hb^2 ( \hb^2 + 8 D \si_0^2 t ) } t \label{eq: deco_func_nd}
\\
\Theta(p, t) \approx \eta \frac{ 4 p_0 \si_0^2 ( p + mg \uptau(t) ) }{ \hb^2 + 8 D \si_0^2 t }   .  \label{eq: phase_nd}
\end{numcases}
In this limit and for times $ t \ll  \si_p^2 / D $, one can introduce the decoherence time defined as 
\begin{eqnarray}
 \tau_D = \frac{ \si_p^4 }{ ( 1 + \eta^2 ) p_0^2 D }; \qquad \Gamma(t) \approx - \frac{t}{\tau_D}
\end{eqnarray}
%
%where we have introduced the decoherence time $ \tau_D $. 
Note that one can get the same result {\it directly} from Eq. (\ref{eq: deco_func}) in this short time limit. 
%This is independent on the relaxation time $1/\ga$.

%
\begin{figure} 
	\centering
	\includegraphics[width=12cm,angle=-0]{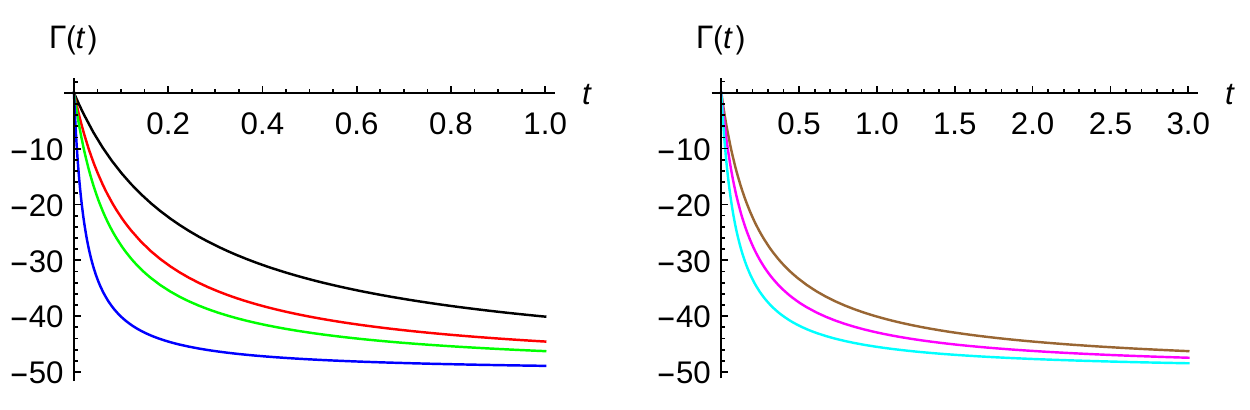}
	\caption{Decoherence function $ \Gamma(t) $ given by (\ref{eq: deco_func}) versus time for $ k_B T = 2 $ (left panel) and for $ \ga = 0.005 $ (right panel). Color curve codes in the left panel are: $ \ga = 0.005 $ (black), $ \ga = 0.01 $ (red), $ \ga = 0.015 $ (green), $ \ga = 0.05 $ (blue); whereas in the right panel, $ k_B T = 2 $ (brown), $ k_B T = 3 $ (magenta) and $ k_B T = 5 $ (cyan). Parameters for the two minimum-uncertainty Gaussian wave packets are $\si_0=5$ and $p_0=-1$. }
	\label{fig: decohere_func} 
\end{figure}

As an illustration, in Fig. \ref{fig: decohere_func} the decoherence function $ \Gamma(t) $ , given by Eq. (\ref{eq: deco_func}), is plotted
versus time for $ k_B T = 2 $ (left panel) and for $ \ga = 0.005 $ (right panel). In the left panel, the curves correspond to $ \ga = 0.005 $ (black), 
$ \ga = 0.01 $ (red), $ \ga = 0.015 $ (green), $ \ga = 0.05 $ (blue).  In the right panel, the curves correspond to $ k_B T = 2 $ (brown), 
$ k_B T = 3 $ (magenta) and $ k_B T = 5 $ (cyan). The initial parameters for the two minimum-uncertainty Gaussian ($\eta= 0$)  wavepackets 
are $\si_0=5$ and $p_0=-1$. In both cases, the asymptotic behavior is reached at relative small times. However, when varying the temperature this
behavior is reached around three times later. In other words, this function decreases faster with $\ga$ than with temperature $k_B T$.

\section{Decoherence for two-identical-particle systems}

%We first derive two-particle CL master equation. 
Eq. (\ref{eq: CLeq_repind}) is linear in $\hat{\rho}$. Writing it as $ \dot{\hat{\rho}} = \hat{\mathcal{L}} \hat{\rho} $, 
$\hat{\mathcal{L}}$ being a linear operator and assuming $ \hat{\rho}_1 $ and $ \hat{\rho}_2 $ are two one-particle states describing 
two non-interacting particles 1 and 2, one can easily sees that the time evolution  for the product state $ \hat{\rho}_1 \otimes \hat{\rho}_2 $ 
is given by
\begin{eqnarray} \label{eq: CL_2p}
\frac{\pa}{\pa t} ( \hat{\rho}_1 \otimes \hat{\rho}_2 ) &=& ( \hat{\mathcal{L}}_1 + \hat{\mathcal{L}}_2 ) ( \hat{\rho}_1 \otimes \hat{\rho}_2 )
\end{eqnarray}
where the linearity of operator $ \hat{\mathcal{L}} $ appearing in the one-particle master equation has to be used. 

Let us  consider now a system of two identical {\it spinless} particles. According to the spin-statistics theorem, the state of such a system must 
have a given symmetry under the exchange of particles; (anti-)symmetric for identical (fermions) bosons.
By taking the initial momentum-space wavefunction as the pure state
\begin{eqnarray} \label{eq: Psi_pm}
\Phi_{\pm}(p_1, p_2, 0) &=& \mathcal{N}_{\pm} \{ \phi(p_1, 0) \chi(p_2, 0) \pm \chi(p_1, 0) \phi(p_2, 0) \}
\end{eqnarray}
$\phi$ and $\chi$ being one-particle wave functions and $\mathcal{N}_{\pm}$ the normalization constant for bosons (+) and fermions (-), 
then the time evolution under the two-particle CL equation (\ref{eq: CL_2p}) yields
\begin{eqnarray} 
\rho_{\pm}(p_1, p_2,p_1', p_2', t) &=& \mathcal{N}_{\pm}^2 \{ \rho_{11}(p_1,p_1', t) \rho_{22}(p_2, p_2', t) 
+ \rho_{22}(p_1, p_1', t) \rho_{11}(p_2, p_2', t) 
\nonumber \\
& & \qquad\pm \rho_{12}(p_1, p_1', t) \rho_{21}(p_2, p_2', t) \pm \rho_{21}(p_1, p_1', t) \rho_{12}(p_2, p_2', t)
\}
\nonumber \\
\label{eq: denmat_2p}
\end{eqnarray}
where
\begin{numcases}~
\rho_{11}(p, p', 0) = \phi_0(p) \phi_0^*(p')
\\
\rho_{22}(p, p', 0) = \chi_0(p) \chi_0^*(p')
\\
\rho_{12}(p, p', 0) = \phi_0(p) \chi_0^*(p')
\\
\rho_{21}(p, p', 0) = \chi_0(p) \phi_0^*(p')   .
\end{numcases}
%
%Diagonal terms $ \rho_{11}(p, p', t) $ and $ \rho_{22}(p, p', t) $ are one-particle densities and off-diagonal %terms $ \rho_{12}(p, p', t) $ and $ \rho_{21}(p, p', t) $ are the so-called coherences. 
Although $ \rho_{11}(p, p', t) $ and $ \rho_{22}(p, p', t) $ are one-particle densities, $ \rho_{12}(p, p', t) $ and $ \rho_{21}(p, p', t) $ are not.
However, all these functions are solutions of one-particle CL equation 
(\ref{eq: CLeq_mom})  satisfying the continuity equation (\ref{eq: con_CL}).
The joint detection probabilities are given by the diagonal elements of Eq. (\ref{eq: denmat_2p}); 
\begin{eqnarray} 
P_{\pm}(p_1, p_2, t) = \mathcal{N}_{\pm}^2 [ P_{11}(p_1, t) P_{22}(p_2, t) 
+ P_{22}(p_1, t) P_{11}(p_2, t) \pm 2\re\{ P_{12}(p_1, t) P_{21}(p_2, t)\} ]
 \label{eq: P_2p}
\end{eqnarray}
where
\begin{eqnarray}
P_{ij}(p, t) &=& \rho_{ij}(p, p, t)
\end{eqnarray}
and the last term of Eq. (\ref{eq: P_2p}) is due to the symmetry of particles. In this context, and due to the environment, this term 
becomes zero along time and we have decoherence in the sense of indistinguishablity loss. 
Note that for distinguishable particles obeying the Maxwell-Boltzmann (MB) statistics, the probability density is given by
\begin{eqnarray} \label{eq: P_2p_MB}
P_{\MB}(p_1, p_2, t) &=& \frac{1}{2} [ P_{11}(p_1, t) P_{22}(p_2, t) + P_{22}(p_1, t) P_{11}(p_2, t) ]  .
\end{eqnarray}
%
%In next section, this effect will be studied by considering one-particle states as cat states corresponding to the {\it two-slit} problem for two identical particles.
%
%
For the single-particle density, $ P_{\sip, \pm}(p, t) = \int_{-\infty}^{\infty} dp_2 \rho_{\pm}(p, p_2; p, p_2, t) $, one obtains
\begin{eqnarray} \label{eq: P_sp}
P_{\sip, \pm}(p, t) &=& \mathcal{N}_{\pm}^2 [  
P_{11}(p, t) + P_{22}(p, t) \pm 2 \re\{ P_{12}(p, t) s(t) \}] 
\end{eqnarray}
where the overlapping integral $s(t)$ is
\begin{eqnarray} \label{eq: st}
s(t) &=& \int_{-\infty}^{\infty} dp' P_{21}(p', t)   .
\end{eqnarray}
%
%In Appendix \ref{app: con-eq} a continuity equation has been derived for the single-particle density.
Due to the continuity equation (\ref{eq: con_CL}), $s(t)$ is a constant which does not depend on  the environment parameters $\ga$ and $T$ and 
time:
$ s(t) = \int dx' P_{21}(p', t) = \int dp' P_{21}(p', 0) = \la \chi(0) | \phi(0) \ra $.
On the other hand, if the system is isolated, states evolve under the Schr\"{o}dinger equation and we have 
\begin{eqnarray} \label{eq: P_sp_Sch}
P_{\sip, \pm}(p, t) &=& \mathcal{N}_{\pm}^2 [  
|\phi(p, t)|^2 + |\chi(p, t)|^2 \pm 2 \re\{ \la \chi(0) | \phi(0) \ra \phi^*(p, t) \chi(p, t) \}]   .
\end{eqnarray}
Comparison of Eq. (\ref{eq: P_sp}) and Eq. (\ref{eq: P_sp_Sch}) reveals that, for open systems, the quantity $ P_{12}(p, t) $ plays the role of $ \phi^*(p, t) \chi(p, t) $. Thus, in analogy to  Eq. (\ref{eq: probden_sup2}), we have again 
\begin{eqnarray}
|P_{12}(p, t)| &=& \sqrt{P_{11}(p, t) P_{22}(p, t)} e^{\Gamma_{12}(t)}
\end{eqnarray}
leading to
\begin{eqnarray} \label{eq: Gamma12}
\Gamma_{12}(t) &=& \log \frac{|P_{12}(p, t)|}{ \sqrt{ P_{11}(p, t) P_{22}(p, t) } }    .
\end{eqnarray}
By considering now one-particle states $\chi$ and $\phi$ as {\it minimum}-uncertainty-product Gaussian wave packets i.e., as in 
Eq. (\ref{eq: wf0}) $\eta=0$, with parameters $ y_0 = 0 $, $q_0$, $\del_0$ and $ x_0 = 0 $, $p_0$, $\si_0$ respectively, one obtains
\begin{eqnarray} 
P_{12}(p, t) &=& \sqrt{ \frac{ 2\si_0  \del_0 }{ \si_0^2 +  \del_0^2 } } \frac{1}{\sqrt{4 \pi b_2(t)}}
\exp \left[ b_0 - \frac{ (p - b_1(t))^2}{ 4 b_2(t) }  \right] \label{eq: P_12}
\\
 s(t) &=& e^{b_0} ~ \sqrt{ \frac{ 2 \si_0  \del_0 }{ \si_0^2 +  \del_0^2 } } \label{eq: st_Gauss}
\\
\mathcal{N}_{\pm} &=& \left \{ 2 \left( 1 \pm \sqrt{ \frac{2\si_0 \del_0}{\si_0^2 + \del_0^2} \exp\left[ - \frac{(p_0-q_0)^2 \si_0^2 \del_0^2}{\hbar^2( \si_0^2 + \del_0^2 )} \right] } \right) \right \}^{-1/2}  \label{eq: normalization}
\end{eqnarray}

where
\begin{eqnarray}
b_0 &=&  - \frac{ \si_0^2 \del_0^2 }{ \si_0^2 + \del_0^2 } \frac{( p_0 - q_0 )^2}{\hb^2}
\label{eq: b0}
\\
b_1(t) &=& e^{ -2\ga t } \frac{ p_0 \si_0^2 + q_0 \del_0^2 }{ \si_0^2 + \del_0^2 } - mg \uptau(t)
 \label{eq: b1}  \\
b_2(t) &=&  e^{ -4\ga t } \frac{ \hb^2 }{ 4 ( \si_0^2 + \del_0^2 ) } + D \frac{ 1 - e^{-4\ga t} }{ 4\ga }. \label{eq: b2}
\end{eqnarray}
Note that for $ \del_0 = \si_0 $ one has $ b_2(t) = w_t^2 / 2 $. $ P_{11}(p, t) $ and $ P_{22}(p, t) $ are 
given by Eq. (\ref{eq: probden}) by using appropriate momenta. 
For $ \del_0 = \si_0 $ from Eq. (\ref{eq: Gamma12}) one obtains 
%, corresponding to the {\it one-slit} diffraction,
%
\begin{eqnarray} \label{eq: Gamma12_Gauss}
\Gamma_{12}(t) &=& - \frac{ \si_0^2 ( p_0 - q_0)^2 }{2\hb^2} \left\{ 1 - \left[ 1 + D \frac{2\si_0^2}{\hb^2 \ga} ( e^{4\ga t} - 1 ) \right]^{-1}
 \right\}  .
\end{eqnarray}
One sees that the decoherence function is negative and the same for both bosons and fermions.
The decoherence process due to the last term of Eq. (\ref{eq: P_sp}) is here interpreted as loss of being indistinguishable as 
described in \cite{MoMi-EPJP-2020-1}.
Notice that the case $ p_0 = q_0 $ can take place only for bosons which then the wave function (\ref{eq: Psi_pm}) takes the product form just as classical states, revealing that quantum statistics is unimportant when the decoherence function $ \Gamma_{12}(t) $ becomes zero.
Another possibility for vanishing the last term of Eq. (\ref{eq: P_sp}) is when the overlapping integral is negligible. In such a case the 
quantum statistics is unimportant too. However, this possibility can also happen in isolated systems and it is not a result of interaction with the 
environment. Therefore, one should consider the effect of environment on $ P_{12}(p, t) $ and $ P_{21}(p, t) $ as an additional source 
of decoherence taking place for identical particle systems.

Decoherence can also be studied through what it is called simultaneous detection probability i.e., measuring the joint detection probability for 
both particles in a given interval of the $p-$space. If we consider a detector, in momentum space, located at the origin with a width $ \Delta $, 
then the {\it ratio} of simultaneous detection probability of indistinguishable particles to the distinguishable ones is given by
\begin{eqnarray} 
p_{\pm}(t) &=& \frac{ p_{\substack{\BE \\ \FD}}(t)
}{ p_{\MB}(t) } = 
	\frac{ \int_{-\Delta/2}^{\Delta/2} dp_1 \int_{-\Delta/2}^{\Delta/2}  dp_2 ~ P_{\pm}(p_1, p_2, t)  }{ \int_{-\Delta/2}^{\Delta/2}  dp_1 \int_{-\Delta/2}^{\Delta/2}  dp_2 ~ P_{\MB}(p_1, p_2, t) } \label{eq: detprob}
\end{eqnarray}
where $\pm$ correspond to bosons (Bose-Einstein statistics) and fermions (Fermi-Dirac statistics), respectively.

\section{Results and discussions}

\begin{figure}
	\centering
	\includegraphics[width=12cm,angle=-0]{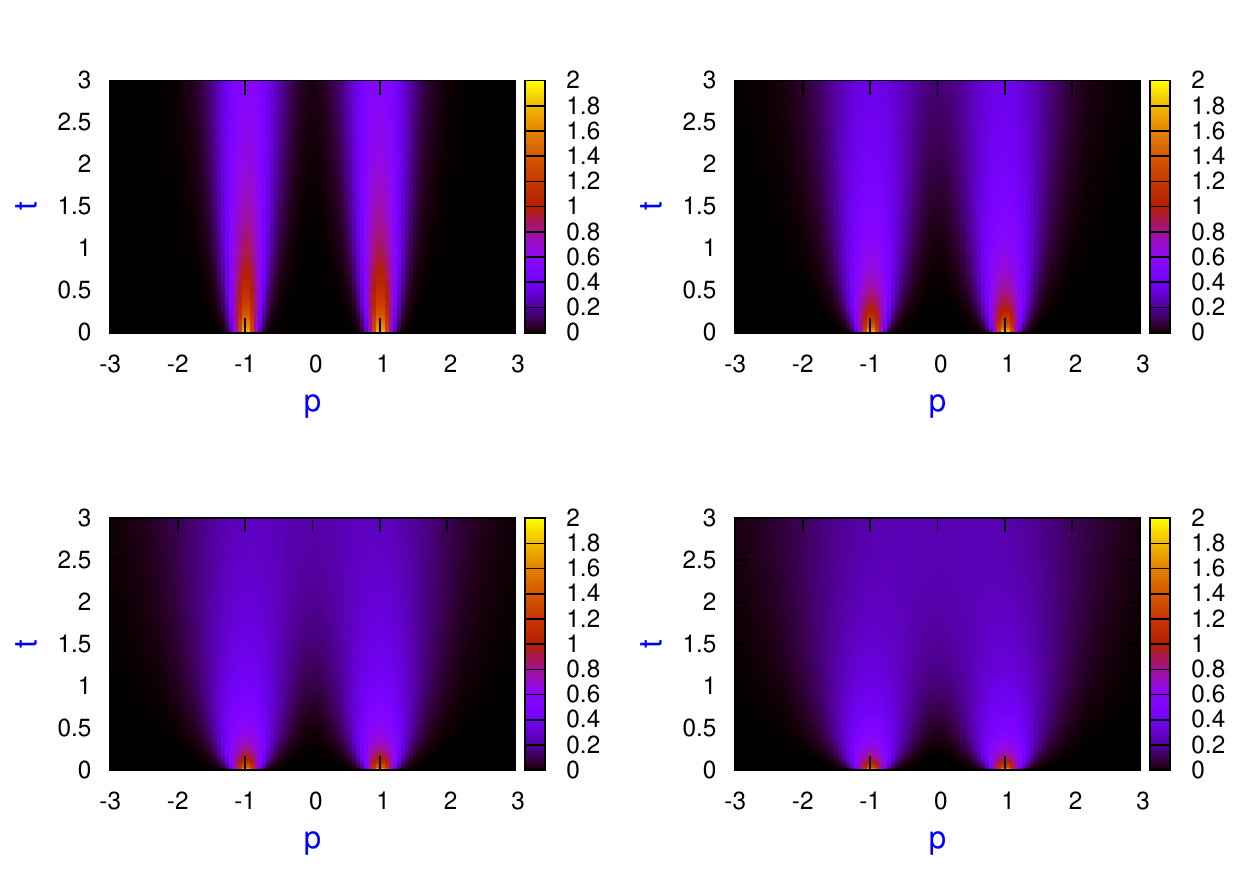}
	\caption{
		Probability density plots (\ref{eq: probden_sup2}) for the superposition of two minimum-uncertainty-Gaussian wave packets, $ \eta = 0 $, for
		$ \ga = 0.005 $ and different values of temperature: $ k_B T = 2 $ (left top panel), $ k_B T = 5 $ (right top panel), $ k_B T = 10 $ (left
		bottom panel) and $ k_B T = 15 $ (right bottom panel). Same parameters as in Figure \ref{fig: decohere_func}.}
	\label{fig: den_sup_Gauss} 
\end{figure}

\begin{figure}
	\centering
	\includegraphics[width=12cm,angle=-0]{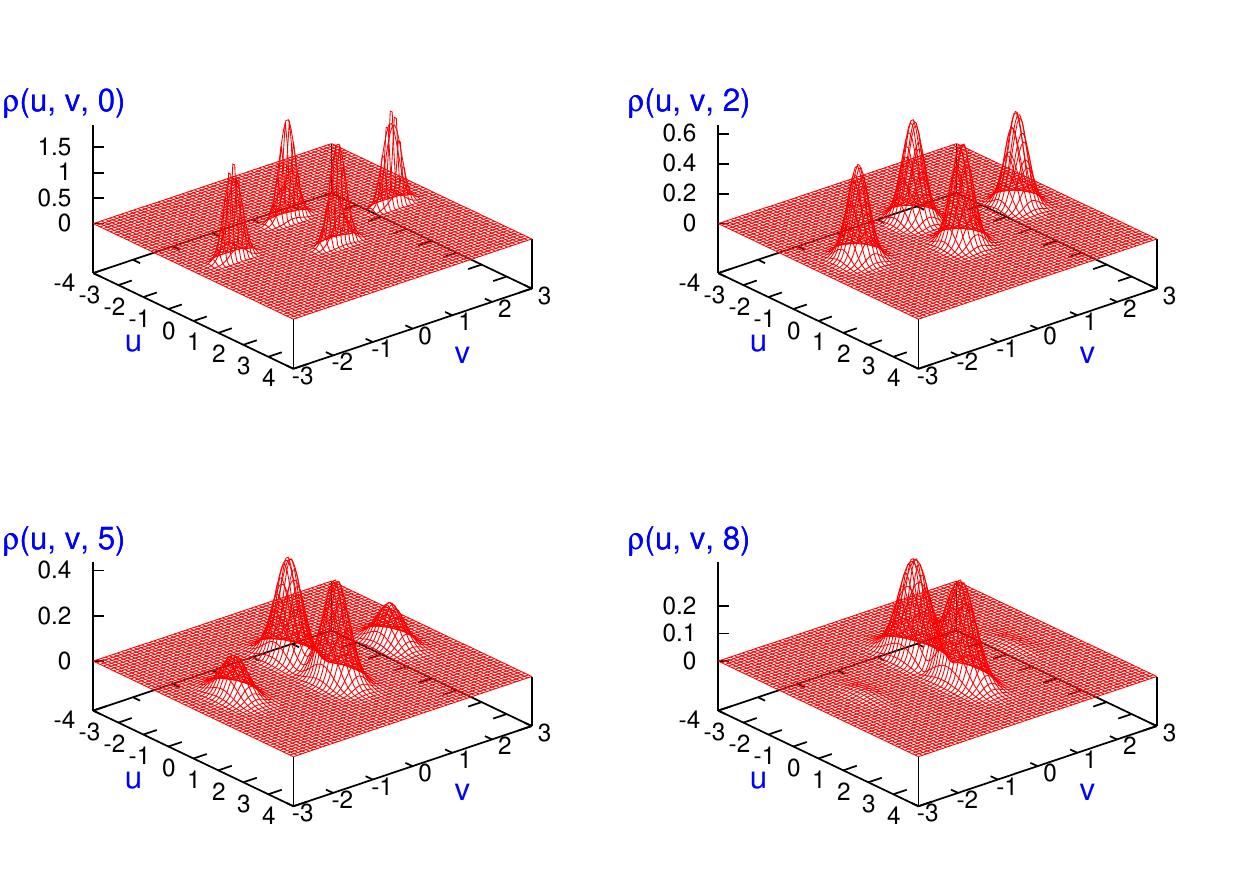}
	\caption{
		Density plots of modulus of density matrix elements given by time evolution of (\ref{eq: rho0}), $ | \rho(u, v, t) | $, in $ uv$-plane at different times $ t= 0 $ (left top panel), $ t = 2 $ (right top panel), $ t = 5 $ (left bottom panel) and $ t = 8 $ (right bottom panel) for $ \eta = 0 $, $ g = 0 $,  $ \ga=0.005 $ and $ k_B T = 2 $. Same parameters as in Figure \ref{fig: decohere_func}. }
	\label{fig: denmat_free_min} 
\end{figure}

Numerical calculations are carried out in a system of units where $ m = \hb = 1 $. In Figure \ref{fig: den_sup_Gauss}, probability densities  (\ref{eq: probden_sup2}) together with 
Eqs. (\ref{eq: deco_func}) and (\ref{eq: phase}) are plotted for the cat state
of two minimum-uncertainty-Gaussian wave packets, $ \eta = 0 $, for $ \ga = 0.005 $ and different values of temperature: $ k_B T = 2 $ 
(left top panel), $ k_B T = 5 $ (right top panel), $ k_B T = 10 $ (left bottom panel) and $ k_B T = 15 $ (right bottom panel). The initial 
parameters used are the same as in Figure \ref{fig: decohere_func}. As discussed previously, no interference pattern is observed in the 
momentum space at any temperature. Obviously, the width of the probability density also increases with time. 
In order to gain some insight on this open dynamics, information about the reduced denrity matrix in the $uv$-plane is helpfull. Thus,
in Figure \ref{fig: denmat_free_min}  density plots  in the $ uv $-plane  at different times are shown for the cat state consisting of two 
minimum-uncertainty-product Gaussian wavepackets in the absence of external potential. The off-diagonal matrix elements $ | \rho(u, v, t) | $ are
shown at $ t= 0 $ (left top panel), $ t = 2 $ (right top panel), $ t = 5 $ (left bottom panel) and $ t = 8 $ (right bottom panel) for $ \eta = 0 $, 
$ g = 0 $,  $ \ga=0.005 $ and $ k_B T = 2 $. It is clearly seen how the coherences or off-diagonal matrix elements goes  to zero at long times. 
The same behavior is observed when the stretching parameter is different from zero as well as the linear potential is present, 
this decoherence process being a little bit faster.
\begin{figure}
\centering
\includegraphics[width=12cm,angle=-0]{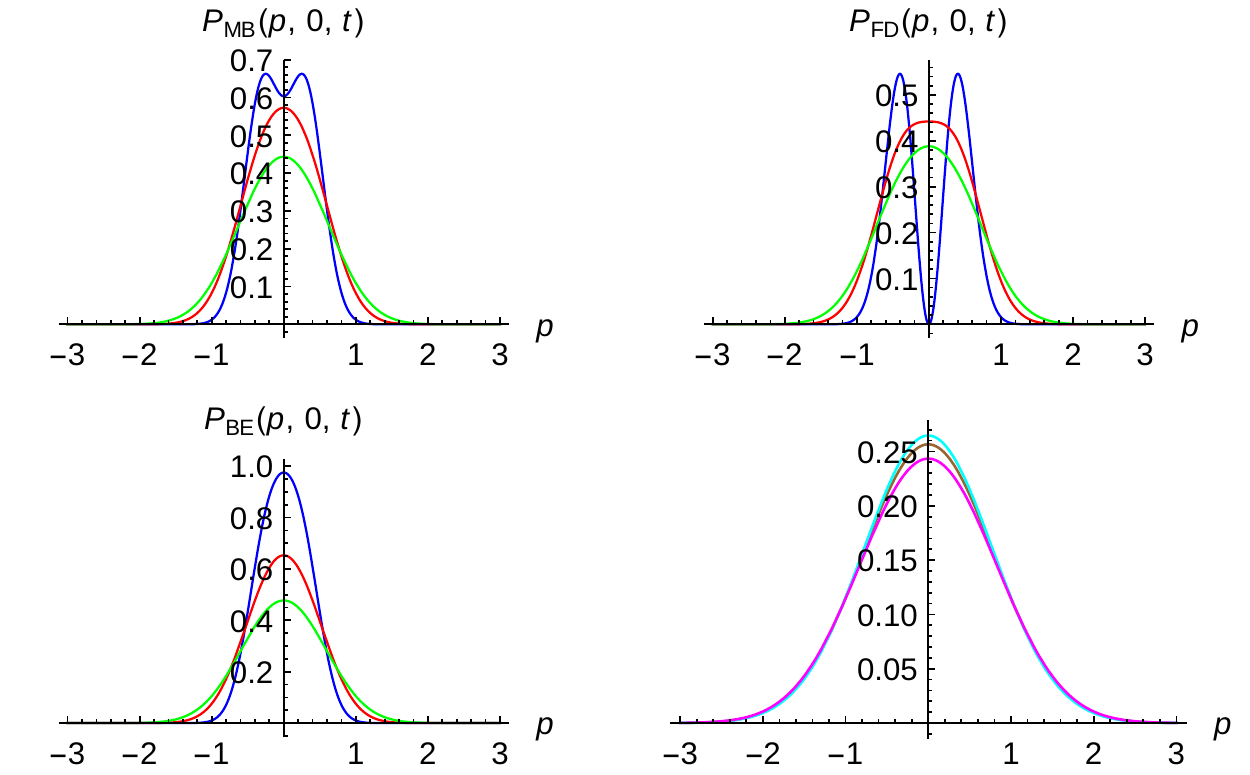}
\caption{
Two-particle probability density plots for finding a particle with zero momentum (\ref{eq: P_2p_MB}) for two distinguishable particles 
obeying MB statistics (left top panel) and (\ref{eq: P_2p}) for two identical bosons (BE statistics, left bottom panel) and fermions 
(FD statistics, right top panel) 
at different times: $ t= 0 $ (blue curves), $ t = 1 $ (red curves) and $ t = 2 $ (green curves). Right bottom panel depicts two-particle 
probability density for distinguishable particles (brown curve), identical bosons (cyan curve) and identical fermions (magenta curve) at 
$ t= 5 $. One-particle states are taken as minimum-uncertainty-Gaussian wavepackets with the same width $ \si_0 = \del_0 = 2 $ and 
opposite kick momenta $ p_0 = - 0.3 $ and $ q_0 = 0.3 $. Environment parameters have been chosen to be $ \ga = 0.005 $ and $ k_B T = 5 $.}
\label{fig: p2p} 
\end{figure}

Decoherence for identical particle systems discussed in previous section can be analyzed in several ways. First, in Figure
\ref{fig: p2p}, two-particle probability density plots for finding a particle with zero momentum and the second one at any value are shown. 
These results are issued from Eq. (\ref{eq: P_2p_MB}) for two distinguishable particles obeying the MB statistics (left top panel) and 
Eq. (\ref{eq: P_2p}) for two identical bosons (left bottom panel) and fermions (right top panel) at different times: $ t= 0 $ (blue curves), 
$ t = 1 $ (red curves) and $ t = 2 $ (green curves). The right bottom panel depicts the same two-particle probability density for distinguishable 
particles (brown curve), 
identical bosons (cyan curve) and identical fermions (magenta curve) at $ t= 5 $. One-particle states are taken as minimum-uncertainty-Gaussian 
wave packets with the same width $ \si_0 = \del_0 = 2 $ and opposite kick momenta $ p_0 = - 0.3 $ and $ q_0 = 0.3 $. The parameters of 
the environment have been chosen to be $ \ga = 0.005 $ and $ k_B T = 5 $. As can be seen, for distinguishable particles obeying the MB 
statistics, the two lobes at $t=0$ describes the two initial separated Gaussian wave packets. With time, the corresponding Gaussian wavepackets 
broaden and the two lobes disappear; the maximum being also around to zero momentum for the second particle. 
For bosons and fermions, the dynamics is quite different. 
The normalization factor also plays an important role since from (\ref{eq: normalization}) one has that $ \mathcal{N}_+ <  \mathcal{N}_{MB} <  \mathcal{N}_- $ where $ \mathcal{N}_{MB} = 1 / \sqrt{2} $. According to 
Eq. \ref{eq: normalization}, for our parameters, $ \mathcal{N}_+ \approx 0.63 $ and $ \mathcal{N}_- \approx 0.81 $. The blue curves 
in each case display different behavior. At the initial time, bosons display a bunching-like behavior and fermions a clear anti-bunching like behavior,
compared to distinguishable particles. For this open dynamics, the last term of Eq. (\ref{eq: P_2p}) together with the overlapping integral 
$s(t)$ governs clearly the time evolution. The decoherence process takes place at $ t \sim 10 $ which is at least  one order of magnitude less 
than the relaxation time $ t_r = 1 / \ga $. It should be emphasized that the decoherence time depends strongly on the
choice of the one-particle states parameters. For instance, for motionless Gaussian wave packets with different widths $ \si_0 = 3 $ and 
$ \del_0 = 0.1 $ where $ \mathcal{N}_+ \approx 0.68 $ and $ \mathcal{N}_- \approx 0.73 $, the decoherence process takes place close to
the relaxation time. With time, the initial bunching and anti-bunching character of the initial distributions are not preserved. Finally, in the right 
bottom panel, the two-particle probability density for the three kind of particles is plotted at $t=5$.
As expected, the time behavior for the three types of particles is quite similar by starting from quite different initial conditions. Notice however that
the fermionic distribution is a little bit broader than the distinguishable and this is similar to the bosonic one. This can be seen as a reminder of the
bunching and anti-bunching character of the initial distributions. 

\begin{figure}
	\centering
	\includegraphics[width=12cm,angle=-0]{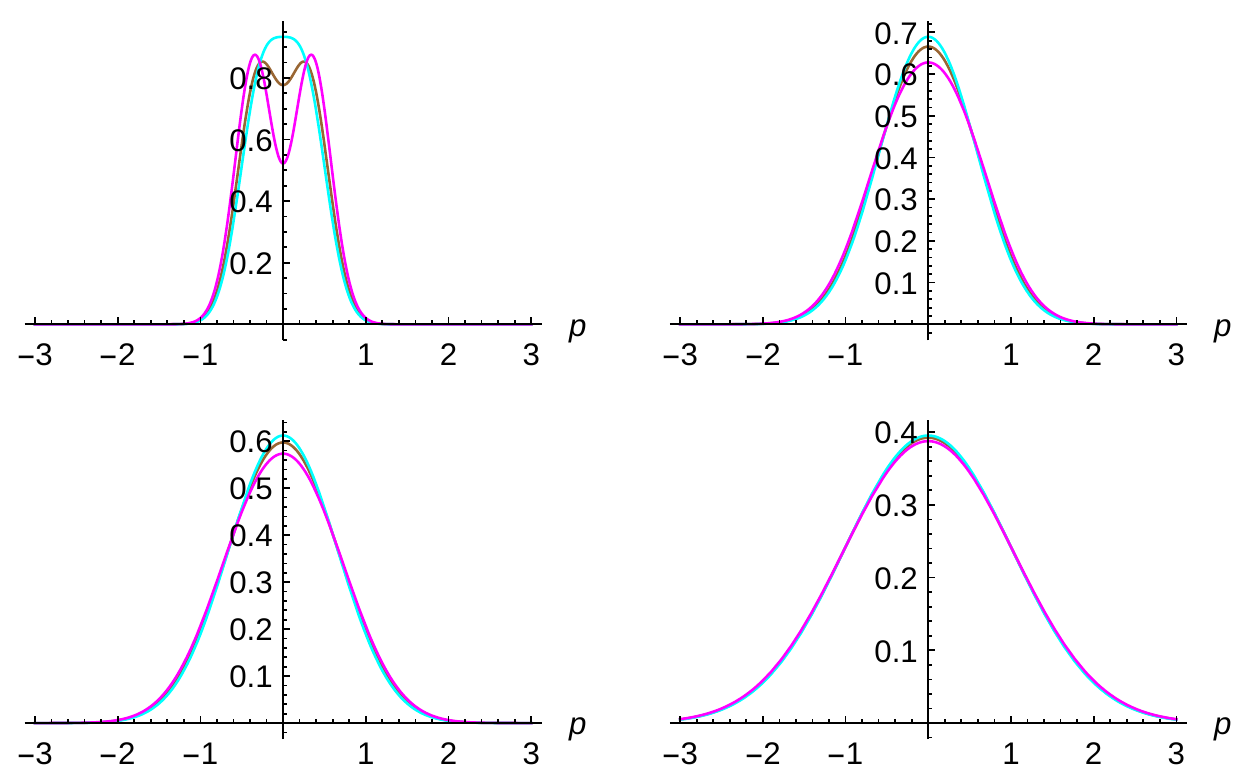}
	\caption{
		Single-particle probability density for distinguishable particles (brown curves); and for identical bosons (cyan curves) and fermions (magenta curves) at different times $ t= 0 $ (left top panel), $ t = 2 $ (right top panel), $ t = 3 $ (left bottom panel) and  $ t = 10 $ (right bottom panel).
		One-particle states are taken as minimum-uncertainty-Gaussian wavepackets with the same widths $ \si_0 = \del_0 = 2 $ but opposite momenta $ p_0 = -0.3 $ and $ q_0 = 0.3 $. Environment parameters have been chosen to be $ \ga = 0.005 $ and $ k_B T = 5 $.}
	\label{fig: psp} 
\end{figure}

The second analysis one can carry out is on the single-patrtile probability. We can ask ourselves which is the corresponding probability density 
for finding a particle with momentum $ p $ independent on the momentum value of the second particle, see Eq. (\ref{eq: P_sp}). 
This is shown in Figure \ref{fig: psp} for distinguishable particles (brown curves), 
identical bosons (cyan curves) and fermions (magenta curves) at different times $ t= 0 $ (left top panel), $ t = 2 $ (right top panel), $ t = 3 $ 
(left bottom panel) and  $ t = 10 $ (right bottom panel). One-particle states are taken as minimum-uncertainty-Gaussian wavepackets with the 
same widths $ \si_0 = \del_0 = 2 $ but opposite momenta $ p_0 = -0.3 $ and $ q_0 = 0.3 $. Environment parameters have been chosen to be 
$ \ga = 0.005 $ and $ k_B T = 5 $. For these parameters the decoherence time is one order of magnitude less than the relaxation time. The same
behavior is observed with respect to the previous figure.

\begin{figure}
	\centering
	\includegraphics[width=8cm,angle=-0]{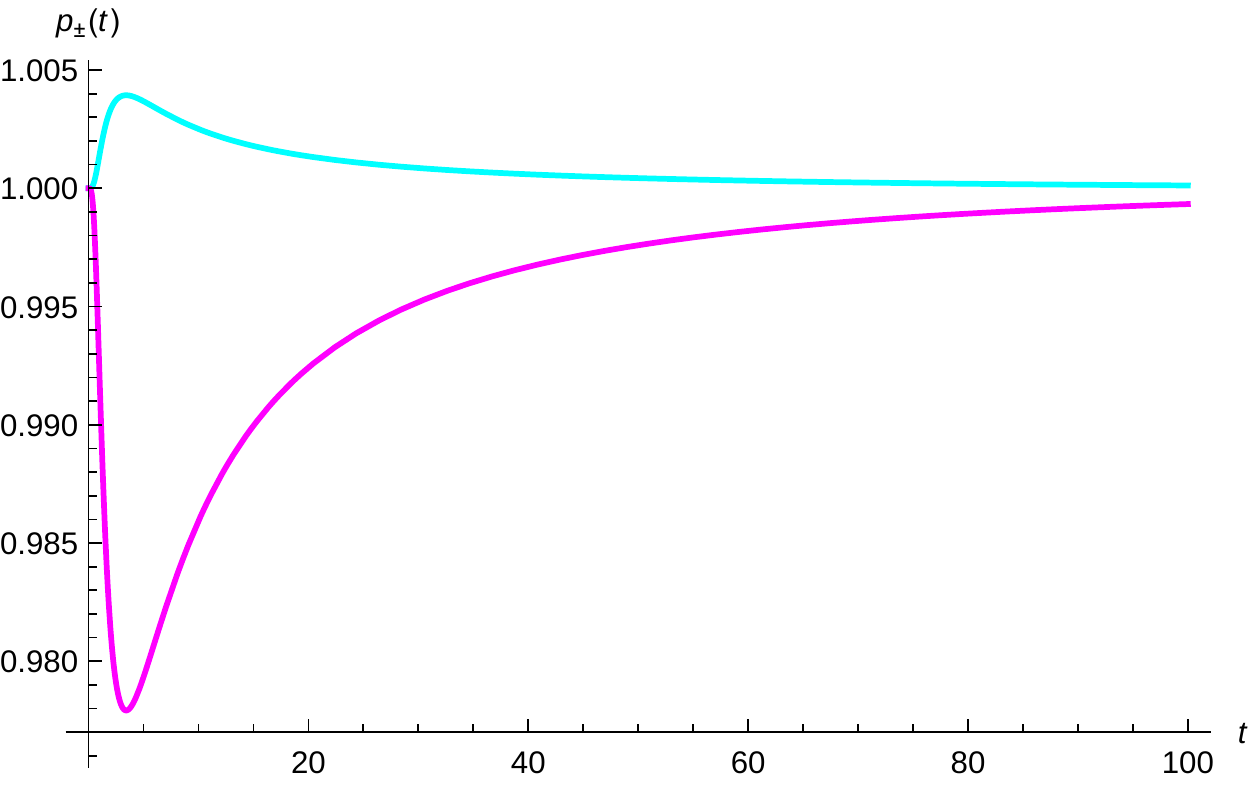}
	\caption{
		Relative simultaneous detection probability $ p_+(t) = \frac{ p_{BE}(t) }{ p_{MB}(t) } $ (cyan) for two identical bosons and $ p_-(t) = \frac{ p_{FD}(t) }{ p_{MB}(t) } $ (magenta) for two identical fermions, measured by a detector with a width $ \Delta = 2 $ located at the origin, see Eq. (\ref{eq: detprob}). One-particle states are taken as minimum-uncertainty-Gaussian wavepackets with the same widths $ \si_0 = \del_0 = 2 $ but opposite momenta $ p_0 = -0.3 $ and $ q_0 = 0.3 $ and the damping constant has been chosen to be $ \ga = 0.005 $.}
	\label{fig: bunch} 
\end{figure}

The third type of analysis is by considering the simultaneous detection probability given by Eq. (\ref{eq: detprob}). In Figure \ref{fig: bunch},
the relative simultaneous detection probability $ p_+(t) = \frac{ p_{BE}(t) }{ p_{MB}(t) } $ (cyan) for two identical bosons and 
$ p_-(t) = \frac{ p_{FD}(t) }{ p_{MB}(t) } $ (magenta) for two identical fermions, measured by a detector with a width $ \Delta = 2 $ 
located at the origin are plotted. One-particle states are taken to be minimum-uncertainty-Gaussian wavepackets with the same widths 
$ \si_0 = \del_0 = 2 $ but opposite momenta, $ p_0 = -0.3 $ and $ q_0 = 0.3 $, and the damping constant has been chosen to be 
$ \ga = 0.005 $. Using Eqs. (\ref{eq: P_2p}) and (\ref{eq: P_2p_MB}) in (\ref{eq: detprob}) yields
\begin{eqnarray}
p_{\pm}(t)	&=& 2 \mathcal{N}_{\pm}^2 \left\{ 1 \pm \frac{ \left| \int_{-\Delta/2}^{\Delta/2}  dp ~ P_{12}(p, t) \right|^2  }
{ \int_{-\Delta/2}^{\Delta/2}  dp ~ P_{11}(p, t) \int_{-\Delta/2}^{\Delta/2} dp ~ P_{22}(p, t) } 
\right\} .
\end{eqnarray}
In this figure, we can  clearly seen that at short times the symmetry of the corresponding wave functions is patent but at asymptotic times
the simultaneous detection probability tends to one; that is, to the classical or MB statistics for both bosons and fermions.

Finally, in this work, we have put in evidence  the quite different behavior of the decoherence process in the momentum space when considering 
cat states and identical spinless particles. Whereas with the first states, no diffraction pattern is observed in this space when comparing with the
configuration space for minimum uncertainty product Gaussian wave packets, a residual manifestation of the well-known bunching and anti-bunching
properties of bosons and fermions is observed with time. This behavior is washed out more rapidly when increasing the damping constant 
and temperature.

\vspace{6pt} 

{\bf Acknowledgments}:
SVM acknowledges support from the University of Qom and SMA  from the Fundaci\'on Humanismo y Ciencia.

%=====================================
% References, variant B: internal bibliography
%=====================================

\end{document}